# Fault-Tolerant Real-Time Streaming with FEC thanks to Capillary Multi-Path Routing

Emin Gabrielyan

Switzernet Sàrl and Swiss Federal Institute of Technology (EPFL)
Lausanne, Switzerland
emin.gabrielyan@{epfl.ch,switzernet.com}

*Abstract* – **Erasure resilient FEC codes in off-line packetized streaming rely on time diversity. This requires unrestricted buffering time at the receiver. In real-time streaming the playback buffering time must be very short. Path diversity is an orthogonal strategy. However, the large number of long paths increases the number of underlying links and consecutively the overall link failure rate. This may increase the overall requirement in redundant FEC packets for combating the link failures. We introduce the Redundancy Overall Requirement (ROR) metric, a routing coefficient specifying the total number of FEC packets required for compensation of all underlying link failures. We present a capillary routing algorithm for constructing layer by layer steadily diversifying multi-path routing patterns. By measuring the ROR coefficients of a dozen of routing layers on hundreds of network samples, we show that the number of required FEC packets decreases substantially when the path diversity is increased by the capillary routing construction algorithm.**

## I. INTRODUCTION

Erasure resilient FEC codes achieve high reliability in off-line streaming in most challenging network conditions [MacKay05], [Shokrollahi04], [Honda04], [Luby02], [Hollywood03]. Off-line streaming can significantly benefit from FEC due to the fact that in contrary to real-time streaming, the receiver is not obliged to deliver in time the "fresh" packets to the user. Since long buffering is not a concern, packets representing the same information can be received at different times.

In real-time single-path streaming, when buffering time is restricted, FEC can only mitigate short granular failures [Johansson02], [Huang05], [Padhye00] and [Altman01]. Packets representing the same information cannot be collected at very remote periods of time. Recent publications show the applicability of FEC in real-time streaming with path diversity. Author of [Qu04] shows that strong FEC improves video communication following two disjoint paths and that in two correlated paths weak FEC is still advantageous. [Tawan04] proposes adaptive multi-path routing for Mobile Ad-Hoc Networks (MANET) addressing the load balance and capacity issues, but mentioning also the possible advantages for FEC. Authors of [Ma03] and [Ma04] suggests replacing in MANET the link level Automatic Repeat Query (ARQ) by a link level FEC assuming regenerating nodes. Authors of [Nguyen02] and [Byers99] studied video streaming from multiple servers. The same author [Nguyen03] later studied real-time streaming over two paths using a static Reed-Solomon RS(30,23) code (FEC blocks carrying 23 source packets and 7 redundant packets). [Nguyen03], similarly to [Qu04] compared two-path scenarios with the single OSPF routing strategy and has shown clear advantages of the double path routing. The path diversity in all these studies is limited to either two (possibly correlated) paths or in the most general case to a sequence of parallel and serial links. Various routing topologies, so far, were not regarded as a ground for searching a FEC efficient pattern.

In this paper we try to present a comparative study for various multi-path routing patterns. Single path routing, being considered as too hostile, is excluded from our comparisons. Steadily diversifying routing patterns are built layer by layer with a *capillary routing* construction algorithm (sections II).

In order to compare multi-path routing patterns, we introduce Redundancy Overall Requirement (ROR) metric, a routing coefficient relying on the sender's transmission rate increases in response to individual link failures. By default, the sender is streaming the media with a static amount of FEC codes in order to tolerate a certain low packet loss rate. The packet loss rate is measured at the receiver and is constantly reported back to the sender with the opposite flow. The sender increases the FEC overhead whenever the packet loss rate is about to exceed the tolerable limit. This end-to-end adaptive FEC mechanism is implemented entirely at the end nodes, at the application level, and is not aware of the underlying routing scheme [Kang05], [Xu00], [Johansson02], [Huang05] and [Padhye00]. The overall number of transmitted adaptive redundant packets for protecting the communication session against link failures is proportional *(1)* to the usual packet transmission rate of the sender, *(2)* to the duration of the communication, *(3)* to the single link failure rate, *(4)* to the single link failure duration and *(5)* to the ROR coefficient of the underlying routing pattern. The novelty brought by ROR is that a routing topology of any complexity can be rated by a single scalar value (section III).

In section IV, we present ROR coefficients of different routing layers built by capillary routing construction algorithm. Network samples are obtained from a random walk MANET with several hundreds of nodes. We show that path diversity achieved by capillary routing algorithm reduces substantially the amount of FEC codes required from the sender.

## II. CAPILLARY ROUTING

Capillary routing may be implemented by an iterative Linear Programming (LP) process transforming a single-path flow into a capillary route. First minimize the maximal value of the load of all links by minimizing an upper bound value applied to all links. The full mass of the flow will be split equally across the possible parallel routes. Find the bottleneck links of the first layer (see below) and fix their load at the found minimum. Minimize similarly the maximal load of all remaining links without the bottleneck links of the first layer. This second iteration further refines the path diversity. Find the bottleneck links of the second layer. Minimize the maximal load of all remaining links, now without the bottlenecks of also the second layer. Repeat this iteration until the entire communication footprint is enclosed in the bottlenecks of the constructed layers.

Fig. 1, Fig. 2 and Fig. 3 show the first three layers of the capillary routing on a small network. The top node on the diagrams is the sender, the bottom node is the receiver and all links are oriented from top to bottom.

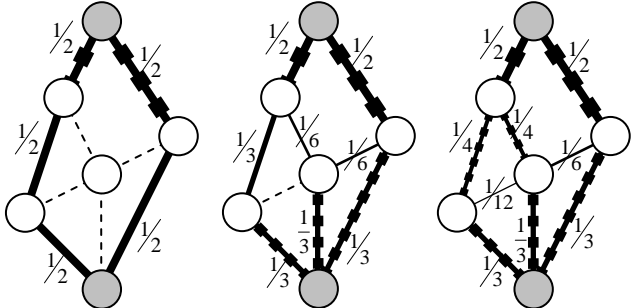

**Fig. 1.** In the first layer the flow is equally split across two paths, two links of which, marked by thick dashes, are the bottlenecks.

**Fig. 2.** The second layer minimizes to 1/3 the maximal load of the remaining seven links and identifies three bottlenecks.

**Fig. 3.** The third layer minimizes to 1/4 the maximal load of the remaining four links and identifies two bottlenecks.

Fig. 4 shows the 10-th layer of capillary routing between a pair of end nodes on a network with 150 nodes and 1364 links. Links not carrying traffic are not shown. The solid lines of the diagram represent the bottleneck links belonging to one of the 10 layers. The dashed lines represent a min-cost solution of the remaining flow not enclosed in bottlenecks after the 10-th layer. There could be several tens of additional routing layers until complete capillarization is achieved.

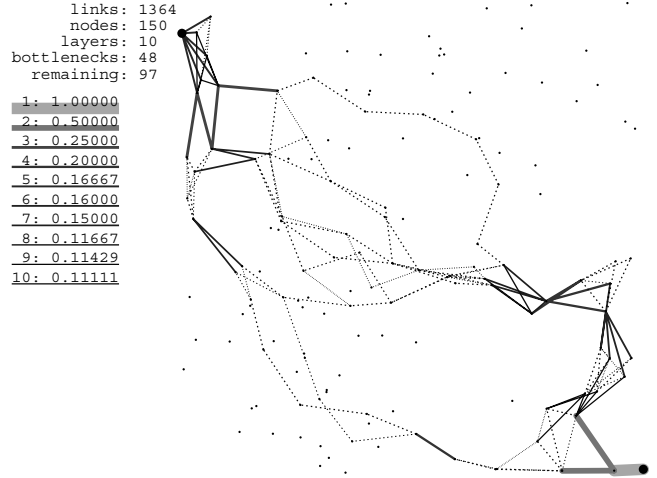

**Fig. 4.** Routing pattern of layer 10 built by capillary routing algorithm on a network sample with 150 nodes

At each layer, after minimizing the maximal load of links, the bottlenecks of the layer are discovered in a bottleneck hunting loop. At each iteration of the hunting loop, we minimize the load of the traffic over all links having maximal load and being suspected as bottlenecks. Links not maintaining their load at the maximum are removed from the suspect list. The bottleneck hunting loop stops if there are no more links to remove.

For capillary routing layers, built simultaneously on 200 independent network samples each with 300 nodes (in average 2,555.7 links per network), Fig. 5 shows the decrease of the number of suspected links during the bottleneck hunting loop of each capillary routing layer from 1 to 10.

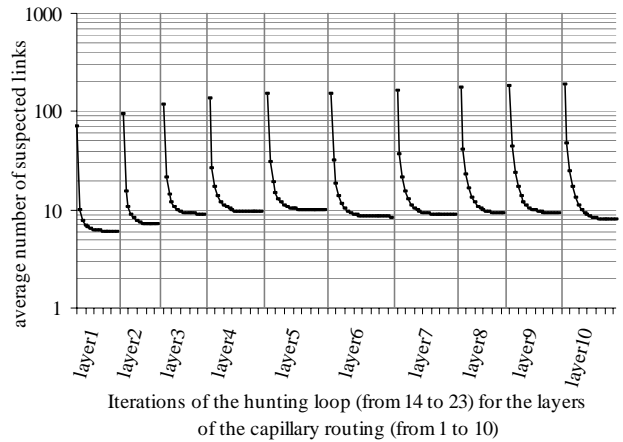

**Fig. 5.** Decrease of the number of suspected links during the bottleneck hunting loop of each of 10 capillary routing layers

At the end of each hunting loop (from 14 to 23 iterations) the suspect list consists of only true bottleneck links, in average between 5.9 and 9.9 bottlenecks per network.

## III. REDUNDANCY OVERALL REQUIREMENT (ROR)

We assume a combination of the little static tolerance of the media stream, combating weak failures, with a dynamically added adaptive FEC combating the strong failures exceeding the tolerable packet loss rate.

For a given routing pattern, ROR is defined as the sum of all transmission rate overheads required from the sender for combating correspondingly all non-simultaneous link failures. For example, if the communication footprint consists of five links, and in response to each individual link failure the sender increases the packet transmission rate by 25%, then ROR will be equal to the sum of these five FEC transmission rate increases, i.e. $ROR = 5 \cdot 25\% = 1.25$. If $P$ is the usual packet transmission rate and $P_l$ is the increased rate of the sender, responding to the failure of a link $l \in L$, where $L$ is the set of all links, then:

$$ROR = \sum_{l \in L} \left( \frac{P_l}{P} - 1 \right) \quad (1)$$

Let us consider a long communication, and let $D$ be the total failure time of a single network link during the whole duration of the communication. $D$ is the product of the average duration of a single link failure, the frequency of a single link failure and the total communication time. According to equation (1):

$$D \cdot P \cdot ROR \;=\; D \cdot P \cdot \sum_{l \in L} \left( \frac{P_l}{P} - 1 \right) \quad (2)$$
$$=\; \sum_{l \in L} (D \cdot P_l - D \cdot P) \quad (3)$$

Assuming a single link failure at a time and a uniform probability and duration of link failures, according to equation (3), $D \cdot P \cdot ROR$ is the number of adaptive redundant packets that the sender actually needs to transmit in order to compensate for all network failures occurring during the total communication time. Therefore ROR is a routing coefficient of the overall number of required redundant packets.

Redundant packets are injected into the original media stream for every block of $M$ source packets. During streaming, $M$ is supposed to stay constant. However, the number of redundant packets for each block of $M$ media packets is variable, depending on the conditions of the erasure channel. The $M$ source packets with their related redundant packets form a FEC block. By $FEC_p$ we denote the FEC block size chosen by the sender in response to a packet loss rate $p$. We assume that by default the media is streamed in FEC blocks of length of $FEC_t$ such that the flow has a static tolerance to weak losses $0 \leq t < 1$. When the loss rate $p$ measured at the receiver is about to exceed the tolerable limit $t$, the sender increases its transmission rate by injecting additional redundant packets.

The random packet loss rate, observed at the receiver during the failure time of a link in the communication path, is the portion of the traffic being still routed toward the faulty link. Thus a complete failure of a link $l$ carrying according to the routing pattern a relative traffic load of $0 \leq r(l) \leq 1$ produces at the receiver a packet loss rate equal to the same relative traffic load $r(l)$.

Equation (1) for ROR can thus be re-written as follows:

$$ROR = \sum_{l \in L \,|\, t \leq r(l) < 1} \left( \frac{FEC_{r(l)}}{FEC_t} - 1 \right) \quad (4)$$

a sum over all links carrying a flow exceeding the tolerable loss limit

The links carrying the entire traffic are skipped in the sum index of equation (4), since the FEC required for the compensation of failures of such links is infinite. By construction (sections II) none of the considered multi-path routing schemes will pass its entire traffic through a non-critical single link.

We compute the $FEC_p$ function assuming a Maximum Distance Separable (MDS) code [Seroussi86], [Schwarz02]. With MDS code we can successfully decode the $M$ source packets if we receive any $M$ packets of the transmission FEC block.

In order to collect a mean of $M$ packets at the receiver under random loss rate $p$, $M/(1-p)$ packets must be transmitted at the sender. However the probability of receiving $M-1$ packets or $M-2$ packets (which makes the decoding impossible) remains high. In order to maintain a very low probability $\delta$ of receiving less than $M$ packets, we must send much more redundant packets in the block than is necessary to receive an average of $M$ packets at the receiver side. We must fix the acceptable Decoding Error Rate (DER), such that $\delta \leq DER$, in order to compute the $FEC_p \geq M$ function.

The probability of having exactly $n$ losses (erasures) in a block of $N$ packets with a random loss probability $p$ is computed according to the binomial distribution:

$\binom{N}{n} \cdot p^n \cdot q^{N-n}$, where $\binom{N}{n} = \dfrac{N!}{n! \cdot (N-n)!}$ and $q = 1 - p$

The probability of having $N - M + 1$ or more losses, i.e. the decoding failure probability, is computed as follows:

$$\delta = \sum_{n = N-M+1}^{N} \binom{N}{n} \cdot p^n \cdot q^{N-n} \quad (5)$$

Therefore for computing the carrier block's minimal length for a satisfactory communication (i.e. $FEC_p$ function), it is sufficient to steadily increase the block length $N$ until the desired decoding error rate (DER) is met.

$FEC_p$ functions divided by $M$ (i.e. transmission rate increase factors $FEC_p/M$) are bounded above by $\log_p(DER)$ when $M=1$ and below by $1/(1-p)$ when $M \to \infty$ (for packet loss rates much larger a very small DER).

The larger the number of media packets $M$ in the FEC block, the smaller the cost of FEC overhead is, but the longer the buffering time at the receiver must be. For example VOIP with 20 ms sampling rate restricts the number of media packets $M$ in a single FEC block to 20 – 25 packets.

If the playback buffering time can be a couple of minutes long, with thousands of source packets in a FEC block (for example in packetized TV) we can assume that $FEC_p = M/(1-p)$. Although for large numbers of source packets MDS codes do not exist, other capacity approaching LDPC [Richardson01] or fountain codes [MacKay05] can decode a large block of source packets requiring only a very little excess of packets (in this context this excess can be ignored).

In such case the overall amount of FEC codes required from the sender as a function of the choice of the multi-path routing pattern, can be evaluated by an ROR coefficient according to the following equation, derived from equation (4) taking into account the above assumptions:

$$ROR = \sum_{l \in L \mid t \leq r(l) < 1} \left( \frac{1-t}{1-r(l)} - 1 \right) \qquad (6)$$

Path diversity can be required in off-line large file downloads aiming at avoiding the idle times of the last kilometer bottleneck occurring due to arbitrary failures elsewhere, within the lossy Internet. Thanks to the sender's adaptive transmission rate and to multi-path routing, one may feed the last kilometer bottleneck link constantly at its maximal bandwidth (see [Nguyen02] and [Byers99] for video streaming from multiple servers). In this case also the choice of the multi-path routing pattern can be rated by equation (6). Note that according to equations (4) and (6) the ROR coefficient of a routing pattern depends also on the static tolerance $t$ of the streaming media to weak failures.

## IV. FEC REQUIREMENT IN CAPILLARY ROUTING

For capillary routing layers 1 to 10, we compute the average ROR coefficients simultaneously over several networks. The network samples are drawn from timeframes of a random walk MANET. Initially the nodes are randomly distributed on a rectangular area, and then, at every timeframe, they move according to a random walk algorithm. If two nodes are close enough (and are within the coverage range) then there is a link between them. At the same time we consider also streaming media at 15 different intensities of static FEC codes tolerating correspondingly weak packet loss rates from 3.6% to 7.8% (with an increment of 0.3%).

Fig. 6, represents a MANET with 115 nodes and 300 timeframes divided into seven sets of network samples. For each set of samples and for each static FEC intensity we plot the average ROR coefficient (over all considered network samples) as the routing layer increases. Fig. 6. shows that the overall requirement in FEC codes decreases with capillarization. The ROR coefficients of the routing samples are computed assuming a short playback buffering time according to equation (4), where the FEC block size (as function of the packet loss rate $p$) is computed according to equation (5), the number of media packets ($M$) per transmission block is 20 and the desired decoding failure rate (DER) is $10^{-5}$.

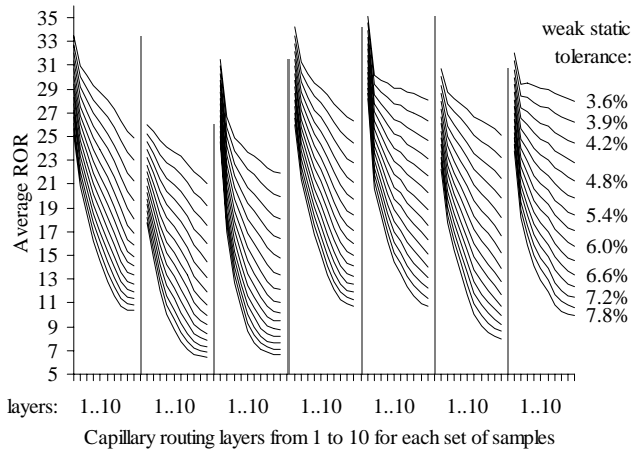

**Fig. 6.** Average ROR as a function from the capillary routing layer

Fig. 7 represents a MANET with 120 nodes and 150 timeframes divided into four sets of network samples. The upper 15 curves similarly to the curves of Fig. 6 are computed according to equations (4) and (5), where $M = 20$ and $DER = 10^{-5}$. However, the lower 15 curves of Fig. 7 are computed according to equation (6) for streaming with large FEC blocks.

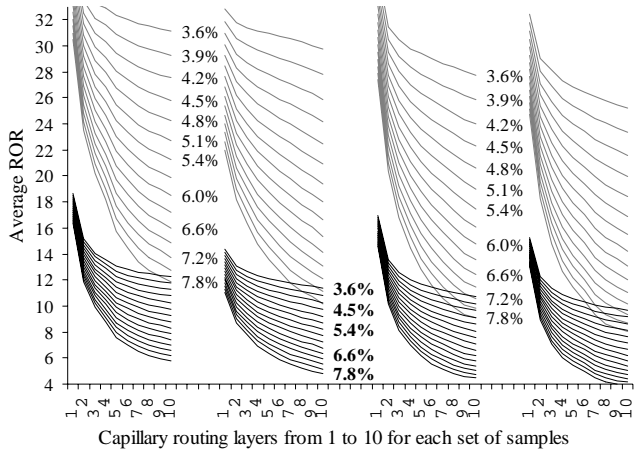

**Fig. 7.** Average ROR computed assuming real-time streaming (the group of curves above) and off-line streaming (the group below)

When streaming with large blocks the Redundancy Overall Requirement is twice as low as in streaming with restricted playback buffering time, but the capillarization of routing is beneficiary in both cases.

Logically, the ROR curve of the media stream is shifted down as the statically added tolerance increases, but the increase of the weak static tolerance yields also a stronger efficiency gain achieved by capillarization. The drawback of path diversity in general is that by forming long paths we may unjustifiably increase the number of links in the communication footprint raising the overall failure rate and thus possibly increasing the overall requirement in FEC codes. Fig. 6 and Fig. 7 however show that despite the communication footprint becomes larger; with the routing patters built by capillary routing algorithm the requirement in redundant packets decreases noticeably most of the time.

## V. Conclusion

The resiliency and reliability issues of packetized real-time streaming are of growing importance. Commercial real-time streaming applications however do not consider channel coding as a serious solution for improving the reliability of communication. That is because in single path communications, even heavy FEC overheads cannot protect against failures lasting more than the short duration of the playback buffer. Recent studies demonstrated that path diversity makes FEC applicable for real-time streaming. By studying a wide range of routing topologies, we show that combination of channel coding with appropriate multi-path routing allows reliable real-time streaming with a low overall requirement in FEC codes.

For this purpose we introduce a layer by layer strategy for building multi-path capillary routing patterns. The first layer provides a simple multi-path solution. As the layer number increases, the underlying routing pattern relies on the network more securely. Unlike max-flow or shortest path solutions, for a given source and destination, by construction (section II) there exists only one solution of capillary routing.

We introduce ROR, a method for rating multi-path routing patterns by a single scalar value. The ROR rating corresponds to the total redundancy overhead that the sending node must provide in order to combat the losses occurring from non-simultaneous failures of links in the communication path. Despite the fact that spreading out of the routing results in the increase of the overall failure rate of underlying links and consecutively maybe also of the need in adaptive FEC codes; with capillarization the overall requirement in FEC codes in fact decreases substantially.

Capillary routing can be applicable to multi-hop mobile wireless networks, where wireless content can be streamed to and from the user via multiple base stations; or to the public internet, where, if the physical routing cannot be accessed, an overlay network can be used [Guven04]. We hope that our investigation will provide some guidelines for future design of path diversity-based real-time streaming systems.

## VI. References

[MacKay05] David J. C. MacKay, "Fountain codes", IEE Communications, Vol. 152 Issue 6, Dec 2005, pp. 1062-1068

[Shokrollahi04] Amin Shokrollahi, "Raptor codes", ISIT'04, June 27 – July 2, page 36

[Honda04] Loring Wirbel, "Deal pushes algorithms into digital radio", April 13, 2004, http://www.commsdesign.com/showArticle.jhtml?articleID=18901216

[Luby02] Michael Luby, "LT codes", FOCS'02, November 16-19, pp. 271-280

[Hollywood03] Mark Fritz, "Digital Dailies Flow Freely from Fountain", April 1, 2003, http://www.emedialive.com/Articles/ReadArticle.aspx?CategoryID=45&ArticleID=5077

[Johansson02] Ingemar Johansson, Tomas Frankkila, Per Synnergren, "Bandwidth efficient AMR operation for VoIP", Speech Coding 2002, Oct 6-9, pp. 150-152

[Huang05] Yicheng Huang, Jari Korhonen, Ye Wang, "Optimization of Source and Channel Coding for Voice Over IP", ICME'05, Jul 06, pp. 173-176

[Padhye00] Chinmay Padhye, Kenneth J. Christensen, Wilfrido Moreno, "A new adaptive FEC loss control algorithm for voice over IP applications", IPCCC'00, Feb 20-22, pp. 307-313

[Altman01] Eitan Altman, Chadi Barakat, Victor M. Ramos, "Queueing analysis of simple FEC schemes for IP telephony", INFOCOM 2001, Vol. 2, Ap 22-26, pp. 796-804

[Qu04] Qi Qu, Ivan V. Bajic, Xusheng Tian, James W. Modestino, "On the effects of path correlation in multi-path video communications using FEC over lossy packet networks", IEEE GLOBECOM'04 Vol. 2, Nov 29 - Dec 3, pp. 977-981

[Tawan04] Tawan Thongpook, "Load balancing of adaptive zone routing in ad hoc networks", TENCON 2004, Vol. B, Nov 21-24, pp. 672-675

[Ma03] Rui Ma, Jacek Ilow, "Reliable multipath routing with fixed delays in MANET using regenerating nodes", LCN'03, Oct 20-24, pp. 719-725

[Ma04] Rui Ma, Jacek Ilow, "Regenerating nodes for real-time transmissions in multi-hop wireless networks", LCN'04, Nov 16-18, pp. 378-384

[Nguyen02] Thinh Nguyen, Avideh Zakhor, "Protocols for distributed video streaming", Image Processing 2002, Vol. 3, Jun 24-28, pp. 185-188

[Byers99] John W. Byers, Michael Luby, Michale Mitzenmacher, "Accessing multiple mirror sites in parallel: using Tornado codes to speed up downloads", INFOCOM 1999, Vol. 1, Mar 21-25, pp. 275-283

[Nguyen03] Thinh Nguyen, P. Mehra, Avideh Zakhor, "Path diversity and bandwidth allocation for multimedia streaming", ICME'03 Vol. 1, Jul 6-9, pp. 663-672

[Kang05] Seong-ryong Kang, Dmitri Loguinov, "Impact of FEC overhead on scalable video streaming", NOSSDAV'05, Jun 12-14, pp. 123-128

[Xu00] Youshi Xu, Tingting Zhang, "An adaptive redundancy technique for wireless indoor multicasting", ISCC 2000, Jul 3-6, pp. 607-614

[Seroussi86] Gadiel Seroussi, Ron M. Roth, On MDS extensions of generalized Reed- Solomon codes, IEEE Transactions on Information Theory, Vol. 32, Issue 3, May 1986, pp. 349-354

[Schwarz02] Thomas S. J. Schwarz, Generalized Reed Solomon codes for erasure correction in SDDS, In Workshop on Distributed Data and Structures, WDAS 2002, Paris, Mar 2002

[Richardson01] Thomas J. Richardson and Rüdiger L Urbanke, Efficient Encoding of Low-Density Parity Check Codes, IEEE Transactions on Information Theory, Vol. 47, No. 2, February 2001, pp. 638-656

[Guven04] Tuna Guven, Chris Kommareddy, Richard J. La, Mark A. Shayman, Bobby Bhattacharjee "Measurement based optimal multi-path routing", INFOCOM 2004, Vol. 1, Mar 7-11, pp. 187-196